\documentclass[1]{elsarticle}

\usepackage{lineno}

\journal{NUCLEAR INSTRUMENTS AND METHODS IN PHYSICS RESEARCH A}

\begin{document}

\begin{frontmatter}
 \title{Simulation of ion behavior in an open three-dimensional Paul trap using a power series method}

 \author[a]{Mustapha Said HERBANE}
 \ead{mherbane@hotmail.com}
 \author[a,b]{Hamid BERRICHE}
 \author[a,c]{Alaa ABD EL-HADY}
 \author[a]{Ghadah AL SHAHRANI}
 \author[d]{Gilles BAN}
 \author[d]{Xavier FLECHARD}
 \author[d]{Etienne LIENARD}

 \address[a]{King Khalid University, Faculty of Science, Department of Physics. P.O. Box 9004.  Abha,Saudi Arabia}
 \address[b]{Laboratoire des Interfaces et Mat\'{e}riaux Avanc\'{e}s, Physics Department-College of Science, University of Monastir. 5019 Monastir-Tunisia}
 \address[c]{Department of Physics, Faculty of Science, Zagazig University. Zagazig 44519, Egypt}
 \address[d]{LPC CAEN-ENSICAEN, 6 Boulevard du Marechal Juin, 14050 Caen Cedex, France}

\begin{abstract}

Simulations of the dynamics of ions trapped in a Paul trap with terms in the potential up to the order 10 have been carried out. The power series method is used to solve numerically the equations of motion of the ions. The stability diagram has been studied and the buffer gas cooling has been implemented by a Monte Carlo method. The dipole excitation was also included. The method has been applied to an existing trap and it has shown good agreement with the experimental results and previous simulations using other methods.

\end{abstract}


\end{frontmatter}


\section{Introduction} \label{sec1}
Paul traps are widely used in different experiments \cite{bibi01}\cite{bibi02}\cite{bibi03}\cite{bibi04}. They are made of few electrodes and the trapping is obtained by the application of a DC and an AC voltage in the radiofrequency domain. Simulations of the trapped ions dynamics are crucial in some experiments such as the experiment of the "Laboratoire de Physique Corpusculaire de Caen (LPC Caen)", where an open three-dimensional Paul trap (the LPCTrap) is used for the determination of the beta-neutrino angular correlation parameter \cite{bibi05}. The error on this parameter depends on the spatial and velocity distributions of the trapped ions. These last have been determined using commercial softwares.

In real radiofrequency Paul traps such as the LPCTrap, the general expression of the potential in spherical coordinates $(\rho,\theta,\varphi)$ is given by \cite{bibi1}\cite{bibi2}
\begin{equation}\label{eq1}
\Phi(\rho,\theta)=(U_{DC}+V_{AC}\cos(\Omega t))\sum_{n=0}^{\infty}b_{n}\rho^{n}P_{n}(cos\theta)
\end{equation}
$U_{DC}$ and $V_{AC}\cos(\Omega t)$ are the DC and AC voltages applied to the trap. $b_{n}$ are constants and $P_{n}$ are the Legendre polynomials of order $n$. Using the cylindrical coordinates $(r,z,\phi)$, with $z$ along the trap's axis, $\rho$ is replaced by $\sqrt{(z^2+r^2 )}$ and $cos\theta=\frac{z}{\rho}$. That transforms expression (\ref{eq1}) into \cite{bibi2}\cite{bibi3}
\begin{equation}\label{eq2}
\Phi(r,z)=(U_{DC}+V_{AC}\cos(\Omega t))\sum_{n=0}^{\infty}C_{n}H_{n}(r,z)
\end{equation}
$C_{n}$ are constants and $H_{n}(r,z)$ are functions of $r$ and $z$.

For the ideal quadrupole trap all the terms of the sum except those for $n=0$ and $n=2$ vanish. The equation of motion for a direction $u = x$, $y$ or $z$ is the Mathieu equation. For every direction the analytical solution exists independently of the 2 other directions \cite{bibi4}. When trapped, an ion performs oscillations at the frequencies
\begin{equation}\label{eq5}
\nu_{u}=(n \pm \frac{\beta_{u}}{2})f
\end{equation}
$f=\frac{\Omega}{2\pi}$ is the frequency of the trapping field, $n$ an integer and $\beta_{u}$ is given by a recursion formula \cite{bibi3}. $\nu_{x}$ and $\nu_{y}$ are equal and are noted $\nu_{r}$. The trapping is possible only for some values of $\beta_u$. This defines the trap's stability diagram. The first region of this diagram corresponds to $\beta_z$ and $\beta_x=\beta_y$ taking the values between 0 and 1.

For real radiofrequency ion traps, terms with n higher than 2 must be included in the expression of the potential (\ref{eq2}). The equations of motion for the 3 directions become coupled and non-linear \cite{bibi2}. When the applied potential has a symmetry along the trap's axis ($z$ axis) and a symmetry about the plane $z=0$, only the even terms exist \cite{bibi5}. By considering the first 6 terms, the potential (\ref{eq2}) becomes \cite{bibi2}\cite{bibi5}\cite{bibi6}
\begin{equation}\label{eq7}
\Phi=(U_{DC}+V_{AC}\cos(\Omega t))(C_{0}+C_{2}H_{2}+C_{4}H_{4}+C_{6}H_{6}+C_{8}H_{8}+C_{10}H_{10})
\end{equation}
With
\begin{eqnarray}\label{eq8}
H_{2} &=& r^{2}-2z^{2}\nonumber \\
H_{4} &=& 8z^{4}-24z^{2} r^{2}+3r^{4}\nonumber \\
H_{6} &=& 16z^{6}-120z^{4} r^{2}+90z^{2} r^{4}-5r^{6}\nonumber \\
H_{8} &=& 128z^{8}-1792z^{6}r^{2}+3360z^{4}r^{4}-1120z^{2}r^{6}+35r^{8}\nonumber \\
H_{10} &=& 256z^{10}-5760z^{8}r^{2}+20160z^{6}r^{4}-16800z^{4}r^{6}+3150z^{2}r^{8}-63r^{10}
\end{eqnarray}
Where $r^{2}$ needs to be replaced by $x^{2}+y^{2}$ when using the cartesian coordinates. The equations of motion for an ion of mass $m$ and charge $Q$ are
\begin{eqnarray}\label{eq13}
d^{2}x/d\xi^{2} &=& T(\xi)x G(x,y,z)\nonumber \\
d^{2}y/d\xi^{2} &=& T(\xi)y G(x,y,z)\nonumber \\
d^{2}z/d\xi^{2} &=& T(\xi)z H(x,y,z)
\end{eqnarray}
Where $\xi=\Omega t$ and
\begin{eqnarray}\label{eq14}
T(\xi) &=& -a_{z}+2q_{z}cos\xi\nonumber \\
G(x,y,z) &=& -\frac{1}{8}-\frac{C_{4}}{16C_{2}}(-48z^{2}+12r^{2})-\frac{C_{6}}{16C_{2}}(-240z^{4}+360z^{2}r^{2}-30r^{4})\nonumber \\
&&-\frac{C_{8}}{16C_{2}}(-3584z^{6}+13440z^{4}r^{2}-6720z^{2}r^{4}+280r^{6})-\frac{C_{10}}{16C_{2}}\nonumber\\
&&(-11520z^{8}+80640z^{6} r^{2}-100800z^{4}r^{4}+25200z^{2}r^{6}-630r^{8} )\nonumber \\
H(x,y,z) &=& \frac{1}{4}-\frac{C_{4}}{16C_{2}}(32z^{2}-48r^{2})-\frac{C_{6}}{16C_{2}}(96z^{4}-480z^{2}r^{2}+180r^{4})\nonumber \\
&&-\frac{C_{8}}{16C_{2}}(1024z^{6}-10752z^{4}r^{2}+13440z^{2}r^{4}-2240r^{6})-\frac{C_{10}}{16C_{2}}\nonumber\\
&&(2560z^{8}-46080z^{6}r^{2}+120960z^{4}r^{4}-67200z^{2}r^{6}+6300r^{8})
\end{eqnarray}
The Mathieu parameters $a_{z}$ and $q_{z}$ are given by
\begin{eqnarray}\label{eq11}
a_{z} &=& -\frac{16QU_{DC}C_{2}}{m\Omega^{2}}\nonumber \\
q_{z} &=& \frac{8QV_{AC}C_{2}}{m\Omega^{2}}
\end{eqnarray}

At King Khalid University in Abha, Saudi Arabia, we are using a radiofrequency Paul trap in order to separate between the different calcium isotopes. It is made of six rings with an axis and a mid-plane of symmetry. In order to realize numerical simulations of our work, we use the SIMION software package \cite{bibi7} which gives accurate results but needs long execution time specially for the systematic study of the  behavior of tens of ions under the action of different trapping voltages. The trap is similar to the LPCTrap \cite{bibi8}. This last has been studied experimentally and with simulation at different occasions \cite{bibi9}\cite{bibi10}\cite{bibi11}\cite{bibi12}. Mainly the ions spatial and velocity distributions, the mean ion's kinetic energy and the oscillation frequencies have been investigated.
In a previous research work, we developed a numerical method to study the behavior of ions in an ideal Paul trap \cite{bibi13}. It is based on the power series solution of the differential equations \cite{bibi14}. It gives accurate results in relatively small calculation time. In this work we present the extension of this method when higher order terms are present in the trapping potential. Our results are compared to the SIMION simulations and to the results of the LPC Caen.

\section{Series solution of the equations of motion} \label{sec2}

We look for the solutions $u(\xi)$ $(u=x,y$ or $z)$ of equation (\ref{eq13}) as power series \cite{bibi14}
\begin{equation}\label{eq15}
u=\sum_{n=0}^{\infty}A_{n}^{u}(\xi-\xi_{0})^{n}
\end{equation}
Where $\xi_{0}$ is an arbitrary constant. The second derivative of $u(\xi)$ is then
\begin{equation}\label{eq16}
\frac{d^{2}u}{d\xi^{2}}=\sum_{n=0}^{\infty}(n+1)(n+2)A_{n+2}^{u}(\xi-\xi_{0})^{n}
\end{equation}
To apply the power series method, we replace the function $T(\xi)$ by its Taylor expansion around $\xi_{0}$.
\begin{equation}\label{eq17}
T(\xi)=\sum_{n=0}^{\infty}A_{n}^{T}(\xi-\xi_{0})^{n}
\end{equation}
The coefficients $A_{n}^{T}$ are calculated as follows
\begin{eqnarray}\label{eq18}
A_{0}^{T} &=& 2q_{z}cos\xi_{0}\nonumber \\
A_{1}^{T} &=& -2q_{z}sin\xi_{0}\nonumber \\
A_{n}^{T} &=& -\frac{A^{T}_{n-2}}{n(n-1)}\ for\ n \geq 2
\end{eqnarray}
$A_{0}^{T}$ is then replaced by $A_{0}^{T}-a_{z}$ to get
\begin{equation}\label{eq19}
A_{0}^{T} = 2q_{z}cos\xi_{0}-a_{z}
\end{equation}
The functions $G$ and $H$ of equation (\ref{eq13}) need also to be replaced by power series. For that we use the coefficients of $x(\xi)$, $y(\xi)$ and $z(\xi)$ to calculate the coefficients of $x^{2}(\xi)$, $y^{2}(\xi)$, $z^{2}(\xi)$. These last are used to calculate the coefficients of $r^2(\xi)$ and those of all the powers and products appearing in $G$ and $H$. We have the property that the product of the series $F_{1}=\sum_{n=0}^{\infty}  A_{n}^{(1)}(\xi-\xi_{0})^{n}$  and $F_{2}=\sum_{n=0}^{\infty}  A_{n}^{(2)}(\xi-\xi_{0})^{n}$ is the series $F_{3}=\sum_{n=0}^{\infty}  A_{n}^{(3)}(\xi-\xi_{0})^{n}$ such that
\begin{equation}\label{eq20}
A_{n}^{(3)} = \sum_{i=0}^{n}A_{i}^{(1)}A_{n-i}^{(2)}
\end{equation}
and $F_{1}+F_{2}$ gives the series $F_{4}=\sum_{n=0}^{\infty}  A_{n}^{(4)}(\xi-\xi_{0})^{n}$ with
\begin{equation}\label{eq21}
A_{n}^{(4)} = A_{n}^{(1)}+A_{n}^{(2)}
\end{equation}
The third property we need to calculate the series of $G$ and $H$ is that if we add a constant term $C$ to a series, $F_{1}$ for example, we get the series $F_{5}=\sum_{n=0}^{\infty}  A_{n}^{(5)}(\xi-\xi_{0})^{n}$ such that
\begin{eqnarray}\label{eq22}
A_{0}^{(5)} &=& C+A_{0}^{(1)}\nonumber \\
A_{n}^{(5)} &=& A_{n}^{(1)}\ for\ n > 0
\end{eqnarray}
So, if we have the series of of $x(\xi)$, $y(\xi)$ and $z(\xi)$ using equations (\ref{eq20}), (\ref{eq21}) and (\ref{eq22}), we find the series of $H$ and $G$ then we calculate the series of $xG$, $yG$ and $zH$. Finally by multiplying these last by the series of $T(\xi)$ given by equation (\ref{eq17}), we get the three second terms of equation (\ref{eq13}). We write them $\sum_{n=0}^{\infty}  S_{n}^{x}(\xi-\xi_{0})^{n}$, $\sum_{n=0}^{\infty}  S_{n}^{y}(\xi-\xi_{0})^{n}$ and $\sum_{n=0}^{\infty}  S_{n}^{z}(\xi-\xi_{0})^{n}$ for the equations of $x$, $y$ and $z$ respectively.

Using equation (\ref{eq16}) and equating both terms of equation (\ref{eq13}) we find the relations
\begin{eqnarray}\label{eq23}
A_{n+2}^{x} &=& \frac{S_{n}^{x}}{(n+1)(n+2)}\nonumber \\
A_{n+2}^{y} &=& \frac{S_{n}^{y}}{(n+1)(n+2)}\nonumber \\
A_{n+2}^{z} &=& \frac{S_{n}^{z}}{(n+1)(n+2)}
\end{eqnarray}
These are recursion relations. In fact, knowing $A_{i}^{x}$, $A_{i}^{y}$ and $A_{i}^{z}$ for the order $i = 0, 1,...,n$, one calculates $S_{n}^{x}$, $S_{n}^{y}$ and $S_{n}^{z}$ then uses equation (\ref{eq23}) to calculate $A_{n+2}^{x}$, $A_{n+2}^{y}$ and $A_{n+2}^{z}$. The coefficients of order 0 (proportional to the initial position) and order 1 (proportional to the initial velocity) are needed to get the others at any order.

Practically, we terminate the sum in equation (\ref{eq15}) at $n = n_{max}$. That means that the coordinates of an ion are given by polynomials of degree $n_{max}$ and not by infinite series. If we want to calculate the trajectory of an ion from $t = 0$ to a relatively large time $t_{max}$, a large value of $n_{max}$ is necessary. However, this will not work in practice because the processor will truncate the numbers when high powers of time  or phase are calculated and gives wrong results. The solution is to divide the large time interval into small intervals of width $\Delta t$ or in an equivalent way, divide the corresponding phase interval into small intervals of width $\Delta \xi$. For every small interval, $\xi_{0}$ of equation (\ref{eq15}) is taken equal to $l\Delta \xi$ with l an integer having the value 0 for the first interval, 1 for the second one and so on. The $n_{max}$ coefficients of the polynomial for $x(\xi)$, $y(\xi)$ and $z(\xi)$ are calculated for every interval. The first 2 of them needed to apply the recursion relations (\ref{eq23}) correspond to
\begin{eqnarray}\label{eq24}
A^{u}_{0} &=& u(\xi_{0})\nonumber \\
A^{u}_{1} &=& \frac{du}{d\xi}|_{\xi=\xi_{0}}=\frac{v_{u}(t=\frac{\xi_{0}}{\Omega})}{\Omega}
\end{eqnarray}
Where $v_{u}$ is the component of the velocity in the $u$ direction.

For the first interval, $A_{0}^{u}$ is the $u$ coordinate of the initial position and $A_{1}^{u}$ is the component of the initial velocity divided by $\Omega$. For the following intervals, $A_{0}^{u}$ and $A_{0}^{u}$ are obtained by imposing the continuity of $u(\xi)$ and its derivative $\frac{du}{d\xi}$.

When we applied this method and solved numerically the equation of motion of an ion in an ideal Paul trap \cite{bibi13}, we found that $n_{max} = 15$ and $\Delta\xi=0.38\pi$, give accurate results with a relatively small calculation time. The position and the velocity of the ion are calculated for $\xi=0,0.38\pi,0.76\pi,...$ means with a time step equal to $19\% $ of the RF period. In the following we use the same values for $n_{max}$ and $\Delta \xi$.

\section{Application to the LPCTrap} \label{sec3}

\begin{figure}[htb]
\begin{center}
\includegraphics[scale=0.8]{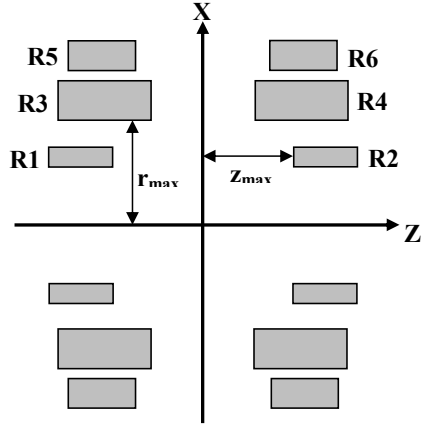}
\caption{Scheme of the LPCTrap. $z_{max} = 8.8\ mm$, $r_{max} = 10.5\ mm$.}
\label{Fig.1}
\end{center}
\end{figure}

The LPCTrap is made of six rings having the same axis of symmetry ($Z$ axis) and it has a median plane of symmetry ($z = 0$ plane). Its general scheme is shown in figure \ref{Fig.1}.

When using the trap, the voltages are applied symmetrically to the rings R1 and R2 while the others are grounded.

We used SIMION7 and calculate the potential for $z$ between $-z_{max} + 0.01\ mm$ and $z_{max} - 0.01\ mm$  and $r$ between 0 and $r_{max}-0.01\ mm$ when $1000\ V$ is applied to R1 and R2 and the others grounded. The potential was calculated in points spaced by $0.01\ mm$ in $z$ and $r$. Fitting the potential by equation (\ref{eq1}) for $U_{DC}=1000\ V$, $V_{AC}=0$ and up to the $100^{th}$ term where we consider only the even terms, gives the constants

\begin{eqnarray}\label{eq25}
C_{0} &=& 0.378473\nonumber \\
C_{2} &=& -3111.505\ m^{-2}\nonumber \\
C_{4} &=& -9.973175\times 10^{4}\ m^{-4}\nonumber \\
C_{6} &=& -1.51631875\times 10^{10}\ m^{-6}\nonumber \\
C_{8} &=& -6.1841484375\times 10^{10}\ m^{-8}\nonumber \\
C_{10} &=& 3.17524609375\times 10^{16}\ m^{-10}
\end{eqnarray}
The fitting is based on the $\chi^{2}$ minimization. $1000 C_{0}=378.473\ V$ is the potential at the trap's center when 1000 V is applied to R1 and R2 and the others grounded. SIMION gives 378.579377 V. When fitting with a degree between 60 and 130,  the maximum variations of these constants relative to their values given in (\ref{eq25}) are
$2.6\times 10^{-4}\ \%$ for $C_{0}$, $4.8\times10^{-4}\ \%$ for $C_{2}$, $3.5\times10^{-2}\ \%$ for $C_{4}$, $9.5\times10^{-3}\ \%$ for $C_{6}$, $9\ \%$ for $C_{8}$ and $0.1\ \%$ for $C_{10}$.

Because $C_{2}$ is negative , the parameter $a_{z}$ is positive if a positive DC voltage is applied to the inner rings and vice versa (equation (\ref{eq11})). $q_{z}$ is always negative. In the experimental work of the LPC Caen, $q_{z}$ was given positive \cite{bibi8}. When using our convention, that means that it is in fact the absolute value of $q_{z}$ which was given.

When simulating the ions behavior in the trap, we considered that an ion is trapped as long as it remains inside the effective trapping volume we define as $|z|<z_{max}$ and $r<r_{max}$. If at some time step, an ion does not fulfill these conditions, it is considered as lost and its trajectory is no longer calculated.

For some of our simulations, we took for the ions uniform random initial positions inside the effective trapping volume. For the velocities we used what is known about the LPCTrap to be as close as possible to the experiment. In fact, the ions kinetic energy distribution before the injection in the trap has been measured \cite{bibi15}. It is a gaussian with the standard deviation $\Delta E = 3\ eV\ FWHM$. We then start by considering ions with random velocities following a Maxwellian distribution with the temperature $T=\frac{\Delta E}{k_{B}}$. $k_{B}$ is the Boltzmann constant. However, most of the ions generated with these initial conditions are lost during the first microseconds of the trapping. We then consider only those which remain trapped after 100 microseconds.

\section{Comparison with SIMION7} \label{sec4}
In order to compare our method to SIMION, we studied the evolution of the position of an ion of mass $40\ amu$ and charge $1\ e$ up to $50\ ms$. We considered separately the cases where the ion was moving in the axial $z$ direction and in the radial $x$ direction. For all the cases we used the frequency of the AC field equal to $1\ MHz$.

\begin{figure}[htb]
\begin{center}
\includegraphics[scale=0.6]{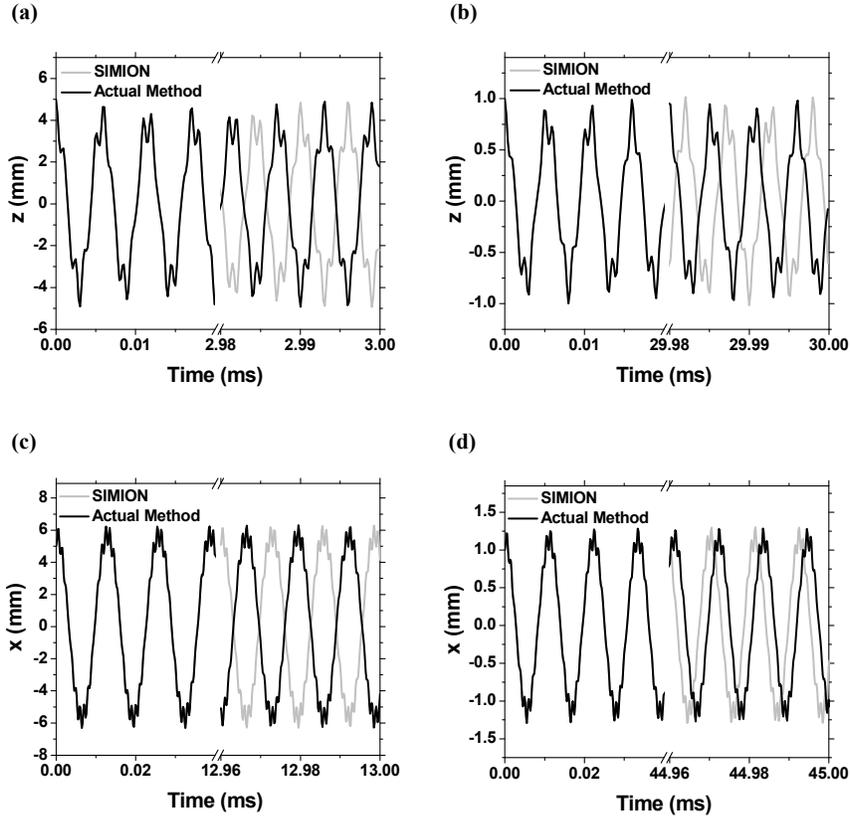}
\caption{Evolution of the position of an ion starting from rest at different initial positions $(x_{0}, y_{0}, z_{0})$. Working point $(q_{z} = -0.5, a_{z} = 0)$ (a) $x_{0}=y_{0}=0, z_{0}=5\ mm$. (b) $x_{0}=y_{0}=0, z_{0}=1\ mm$. (c) $x_{0}=5\ mm, y_{0}=z_{0} =0$. (c) $x_{0}=1\ mm, y_{0}=z_{0} =0$.}
\label{Fig.2}
\end{center}
\end{figure}

Figure \ref{Fig.2} shows the positions for $V_{AC} = 329\ V$ and $U_{DC}=0$ which corresponds to ($q_{z} = -0.5$, $a_{z} = 0$) calculated using equation (\ref{eq11}).

There is a good overlap between the positions calculated by SIMION and those with our method for the first periods of the ion's motion. Figure \ref{Fig.2} (a) corresponds to the an ion starting from rest at $t=0$ and $z=5\ mm$, $x=y=0$. The overlap between the 2 curves ceases for long times. There is a phase shift between the 2 curves which increases with time and is equal to about $\pi$ around $t=3\ ms$. Figure \ref{Fig.2} (b) shows the curve in the same conditions but with the ion starting from $z=1 mm$. The agreement between the 2 curves is good for longer times and the phase shift is less than $\pi$ for $t=30\ ms$. Figure \ref{Fig.2} (c) and (d) gives the same simulations but with an ion starting from rest at $x=5\ mm$ and $x=1\ mm$ respectively with $y$ and $z$ equal to 0. Again a phase shift is observed between our results and those of SIMION. Similar to the motion in the $z$-direction, the rate of the increase of this phase shift is the smaller for the smallest oscillation amplitude. This can be explained by the fact that when the excursion of the ion is large, it enters in regions of space where terms in the potential of order higher than 10, and which we do not include in our simulations, become important. These terms exist in SIMION and they give the difference between the results of the 2 methods (SIMION and our method).

\begin{figure}[htb]
\begin{center}
\includegraphics[scale=0.6]{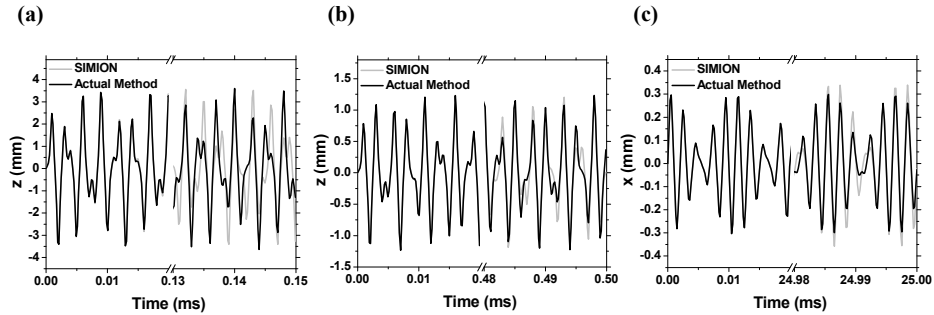}
\caption{Evolution of the position of an ion starting from the origin with different initial kinetic energies $E_{0}$. Working point $(q_{z} = -1.32, a_{z} = -0.54)$ \textbf{(a)} $E_{0}=0.1\ eV$ with the motion in the z-direction. \textbf{(b)} $E_{0}=0.01\ eV$ with the motion in the z-direction. \textbf{(c)} $E_{0}=0.1\ eV$ with the motion in the x-direction.}
\label{Fig.3}
\end{center}
\end{figure}

Figure \ref{Fig.3} shows other examples corresponding to $V_{AC} = 870\ V$ and $U_{DC}=-178\ V$ that is $(q_{z} = -1.32, a_{z} = -0.54)$. This working point lies in the limits of the stability diagram with the $\beta_{x}=\beta_{x}$ and $\beta_{z}$ close to unity. The ion starts from the origin at $t=0$ and with initial kinetic energy of $0.1$ and $0.01\ eV$ for the oscillation in the $z$-direction and $0.1\ eV$ for the $x$-direction. The difference between our results and those of SIMION are seen faster than for the cases of Figure \ref{Fig.2}. We also observe that the agreement between the curves of SIMION and ours depend on the amplitudes of oscillation. The smaller they are the better it is.

In all cases, the amplitudes and the frequencies of oscillation we get, are close to those obtained by SIMION. The main advantage of our method is that it allows the calculation of an ion's trajectory in relatively short time. Using a DELL computer of $3.1\ GHz$ frequency, we get the positions and velocities up to $50\ ms$ in less than $4\ s$ while several minutes are needed by SIMION7. Our codes are written in the C language under Windows and SIMION was used in its default computational quality.

Since we have similar oscillation amplitudes and frequencies when using our method and the SIMION software, we expect to get similar results with the 2 methods when considering the maximum excursion of the ions and their frequencies. The observed phase shift does not have an important effect.

\section{The stability Diagram} \label{sec5}
In order to get the first stability domain of the trap, we start by generating 500 ions of mass $40\ amu$ and charge $1\ e$. These are $Ca^{+}$ ions. We take for them uniform random positions with $-z_{max}<z<z_{max}$ and $r<r_{max}$. The velocities are those of a maxwellian distribution with a temperature of $1000\ K$. These can be ions created by the ionization of hot atoms. For every working point, the trajectories are calculated up to a trapping time of $2\ ms$. At the end, the number of ions $N(q_{z},a_{z})$ which remain in the trap is counted. These are those which satisfy the trapping condition given in section \ref{sec3}. The RF frequency is $1\ MHz$.

\begin{figure}[htb]
\begin{center}
\includegraphics[scale=0.7]{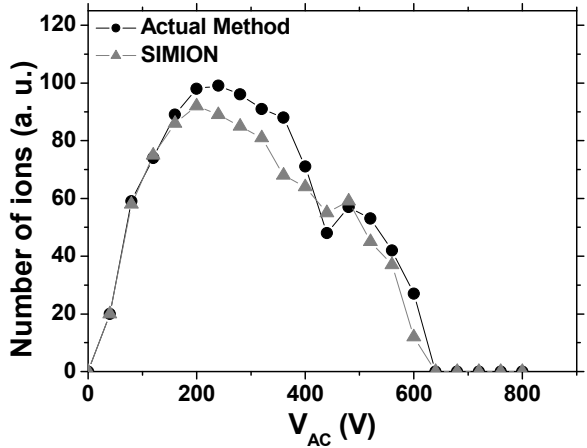}
\caption{Number of trapped ions as function of $V_{AC}$ for $U_{DC}=0$}
\label{Fig.4}
\end{center}
\end{figure}

Figure \ref{Fig.4} gives the result when $V_{AC}$ is scanned between 0 and $800\ V$ with a step $\Delta V_{AC}$ of $40\ V$. $U_{DC}$ is taken 0. That is a scan of $q_{z}$ between 0 and -1.2167 with a step $q_{z}=-0.06$ and $a_{z}=0$.  Our result are compared to those we get when using SIMION. We see that the 2 curves have the same general shape. Specially there is a relative minimum for both of them at the same value of $V_{AC}=440\ V$ ($q_{z}=-0.67$).

\begin{figure}[htb]
\begin{center}
\includegraphics[scale=0.7]{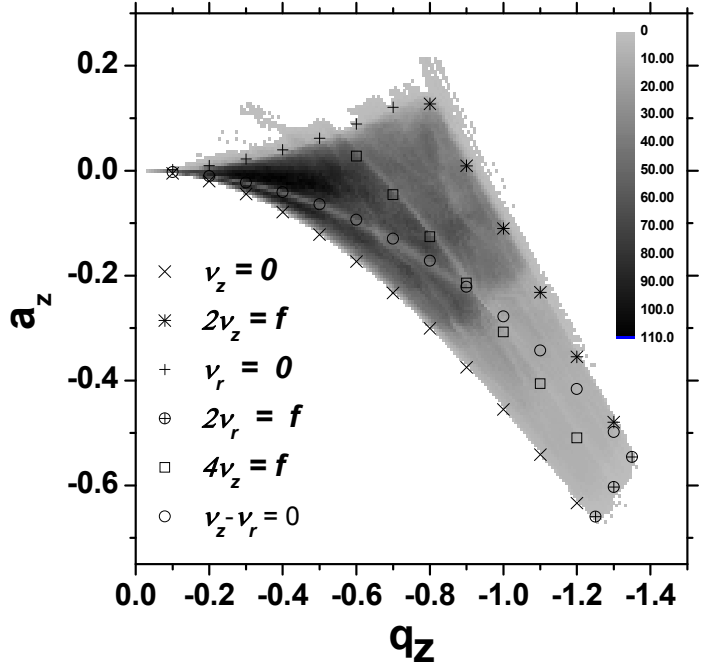}
\caption{The first stability domain of the LPCTrap as obtained by the simulations for $Ca^{+}$ ions. The symbols give the theoretical limits of the domain and the positions of the non linear resonances.}
\label{Fig.5}
\end{center}
\end{figure}

In figure \ref{Fig.5}, we show a more detailed investigation of the trap's stability diagram. For that we scanned $V_{AC}$ between 0 and $1000\ V$ with a step $\Delta V_{AC}=5\ V$. That is $q_{z}$ between 0 and -1.52 with $\Delta q_{z}=-0.0076$. For every value of $V_{AC}$, $U_{DC}$ is scanned between $-230\ V$  and $80\ V$ with $2\ V$ step which corresponds to $a_{z}$ between -0.7 and 0.21 with $\Delta a_{z}=0.006$.
Figure \ref{Fig.5} shows a 2D plot of $N(q_{z},a_{z})$ as a gray scale map. One recognizes the classical shape of the first stability domain. In the domain we see lines where the number of ions is reduced. They are the non linear resonances \cite{bibi2} which occur when
\begin{equation}\label{eq26}
n_{z}\nu_{z}+n_{r}\nu_{r} = nf
\end{equation}
$n_{z}$, $n_{r}$ and $n$ are integers.

A non linear resonance is induced by the terms of the trapping potential having an order larger than 2. When it occurs, the ions oscillation amplitudes become larger and some of them can be lost.

We identify the resonances $4\nu_{z}=f$ and $\nu_{z}-\nu_{r}=0$. This is confirmed by the analytical calculation of these resonances obtained using the recursion relation to get $\nu_{z}$ and $\nu_{r}$ \cite{bibi3}. The analytical results are shown by symbols in the figure. The line $4\nu_{z}=f$ is induced by the absorption by the ions of energy from the RF field and has been observed in some experiments with real traps \cite{bibi021}\cite{bibi022}. The line $\nu_{z}-\nu_{r}=0$ corresponds to the coupling between the ions different degrees of freedom and has also been observed experimentally \cite{bibi023}.

\section{The buffer gas cooling}\label{sec6}

The collision of the trapped ions with the atoms or the molecules of a buffer gas induces their cooling or heating \cite{bibi16}. When the mass of the ion is much bigger than the mass of the buffer gas atoms, its energy is reduced (cooling). The Monte Carlo simulation of this phenomenon has been implemented since decades \cite{bibi6}. The most simple method is based on the Langevin collision theory where the collision probability is independent on the ion's velocity. The other method is the hard sphere model (HS1) which considers the ions and the gas particles as spheres of radii $r_{i}$ and $r_{g}$ respectively. The collision cross section is given by
\begin{equation}\label{eq27}
\sigma_{hs}=\pi r_{hs}^{2}=\pi (r_{i}+r_{g})^{2}
\end{equation}
Its implementation is more complicated than the Langevin theory but it has been shown that the convenience of a model or the other depends on the parameter \cite{bibi17}
\begin{equation}\label{eq28}
\epsilon=\frac{e^{2}\alpha_{e}}{2(4\pi \varepsilon_{0})^{2}k_{B}T_{g}r_{hs}^{4}}
\end{equation}
$\alpha_{e}$ is the gas particle's electric polarizability, $\varepsilon_0$ the electric permittivity of vacuum and $T_{g}$ is the gas temperature. If $\epsilon \ll 1$, the hard sphere collisions dominate. In our case we want to study the cooling of $^{6}Li^{+}$ ions by molecular Hydrogen at room temperature ($300\ K$). Using $\alpha_{e}=0.8\ \AA^{3}$ \cite{bibi18} and $r_{hs}=5.3\ \AA$ (this value will be justified bellow), the parameter $\epsilon$ is about 0.27. We then choosed the model HS1.

In the HS1 model, the collision probability per unit time can be calculated by \cite{bibi19}
\begin{equation}\label{eq29}
\frac{dP}{dt}=n\sigma_{hs}v
\end{equation}
$n=\frac{p}{k_{B} T_{g}}$ is the buffer gas density, $v$ is the mean ion's velocity relative to the buffer gas particles and $p$ the buffer gas pressure. When considering a Maxwellian distribution for the gas velocities, $v$ has the approximate expression \cite{bibi20}
\begin{equation}\label{eq30}
v=\sqrt{v_{i}^2+\overline{v_{g}}^{2}}
\end{equation}
Where $v_{i}$ is the ion's speed, $\overline{v_{g}}=\sqrt{\frac{8k_{B}T_{g}}{\pi m_{g}}}$ is the mean speed of the gas particles and $m_{g}$ their mass.

The collisions between an ion and a buffer gas particle are supposed to be elastic. To get the ion's velocity after the collision, a reference frame other than the laboratory frame is more convenient. SIMION8 uses a frame where the gas particle is stationary \cite{bibi21}. Parks and Szoke use the center of mass (CM) frame \cite{bibi22}. We follow their procedure. To get the velocity of the ion after the collision, its velocity and the velocity of the buffer gas particle in the CM frame before the collision are calculated. The z-axis is rotated and made parallel to the ion's velocity. The collision then occurs in the z direction. The angle $\alpha$ between the line connecting the centers of the 2 colliding spheres and the z-axis defines the impact parameter. It varies between 0 and $\frac{\pi}{2}$. The orientation of the 2 spheres about the rotated z-axis is defined by an angle $\varphi_{r}$ taking the values between 0 and $2\pi$. Knowing these angles, the velocity of the ion after the collision is calculated in the rotated CM frame then transformed back to the laboratory frame.

For every ion, the collision probability is calculated every time step, and a random number between 0 and 1 is generated. If this number is smaller or equal to the calculated probability, the collision is supposed to occur. A random velocity with components following maxwellian distributions with the temperature $T_{g}$ are affected to the buffer gas particle. The new velocity of the ion is calculated according to the procedure explained above.Two additional random numbers are needed. One between 0 and 1 whose value is affected to $sin^{2}\alpha$, and a second between 0 and $2\pi$ affected to $\varphi_{r}$ \cite{bibi22}.

In the LPC Caen experiment, the main buffer gas is molecular hydrogen. The cooling of singly charged ions of mass $6\ amu$ has been simulated for a trapping frequency of $1.3\ MHz$ and $V_{AC} = 80\ V$ \cite{bibi9} with a buffer gas pressure of $5\times 10^{-4}\ mbar$. The same reference gives the experimental cooling times $\tau$ of $^{6}Li^{+}$ for the pressures of $6\times 10^{-6}$, $10^{-5}$ and $4.3\times 10^{-5}\ mbar$. They are 12.9, 7.2 and $2.5\ ms$ respectively. The final temperature of the ions has also been measured by applying electric pulses to the rings R1, R2 and R3 and measuring the time of flight (TOF) to a microchannel plate detector \cite{bibi10}. The comparison of the TOF to SIMION8 simulations give an ion's mean kinetic energy of $0.11\ eV$.

In our simulations, the test of the occurrence of a collision is realized every time step. When a collision occurs, the velocity given by equation (\ref{eq24}) is changed according to the described procedure. For every working point, the initial positions and velocities of the ions were generated according to the procedure described in section \ref{sec3}. We first studied the evolution of 1000 $^{6}Li^{+}$ up to $90\ ms$. For that, we start by fixing a time step for the calculation of some physical quantities of the ion cloud. The latter are the standard deviations $\sigma_{x}$, $\sigma_{y}$, $\sigma_{z}$, $\sigma_{v_{x}}$, $\sigma_{v_{y}}$ and $\sigma_{v_{z}}$ for the positions and the velocities and the ion's mean kinetic energy $\langle E \rangle$. They are calculated every simulation time step during the first 5 periods of the trapping field following every calculation step. We then calculate the mean values of the 7 quantities for the 5 periods. We get $\langle\sigma_{x}\rangle$, $\langle\sigma_{y}\rangle$, $\langle\sigma_{z}\rangle$, $\langle \sigma_{v_{x}} \rangle$, $\langle \sigma_{v_{y}} \rangle$, $\langle \sigma_{v_{z}} \rangle$ and $\langle E \rangle$ for every calculation step.

\begin{figure}[htb]
\begin{center}
\includegraphics[scale=0.6]{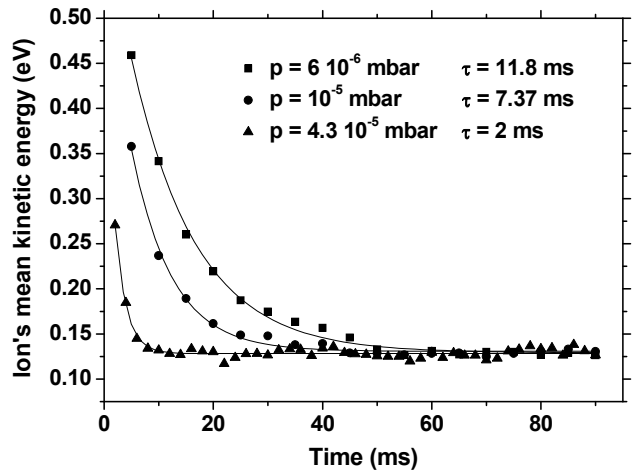}
\caption{Cooling times for Ca ions in a $H_{2}$ buffer gas at three different pressures.}
\label{Fig.6}
\end{center}
\end{figure}

We varied the value of the collision cross section $\sigma_{hs}$ until getting the closest cooling time to the experimental value for the pressure of $10^{-5}\ mbar$. $\tau$ is obtained by fitting the ion's mean kinetic energy by a function exponentially decaying in time. We fixed for that $U_{DC}$  to 0, $V_{AC}$  to $65\ V$ and $f$ to $1.15\ MHz$. These are the usual working parameters in the LPC Caen \cite{bibi8}. We get $\sigma_{hs}=90\ \AA^{2}$, corresponding to a cooling time of $7.37\ ms$. This is shown in figure \ref{Fig.6}. When used with the pressures of $6\times 10^{-6}$ and $4.3\times 10^{-5}\ mbar$, we find cooling times of $11.8$ and $2\ ms$ respectively, which are also close to the experimental values.

The figure shows that after a trapping time of about $55\ ms$, the ions are thermalized. The average value of the mean kinetic energy considered for the points after thermalization is $0.128\ eV$ for $6\times 10^{-6}\ mbar$, $0.129\ eV$ for $10^{-5}\ mbar$ and $0.128\ eV$ for $4.3\times 10^{-5}\ mbar$. This is in good agreement with the experimental results \cite{bibi10}.

From $\sigma_{hs}=90\ \AA^{2}$ we calculate $r_{hs}=5.3\ {\AA}$. this last gives $\epsilon =\ 0.27$ (equation (\ref{eq28})) which justifies the use of the hard sphere model.

\begin{figure}[htb]
\begin{center}
\includegraphics[scale=0.5]{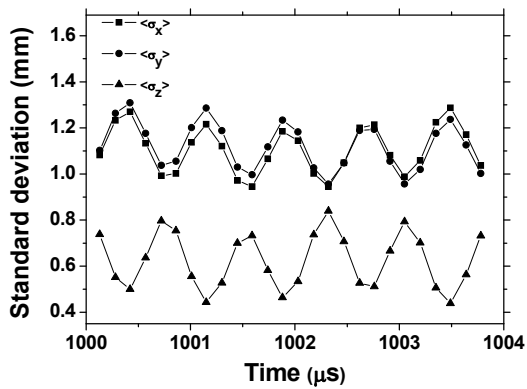}
\caption{Standard deviation of the positions around a trapping time of $1000\ \mu s$.}
\label{Fig.7}
\end{center}
\end{figure}

\begin{figure}[htb]
\begin{center}
\includegraphics[scale=0.5]{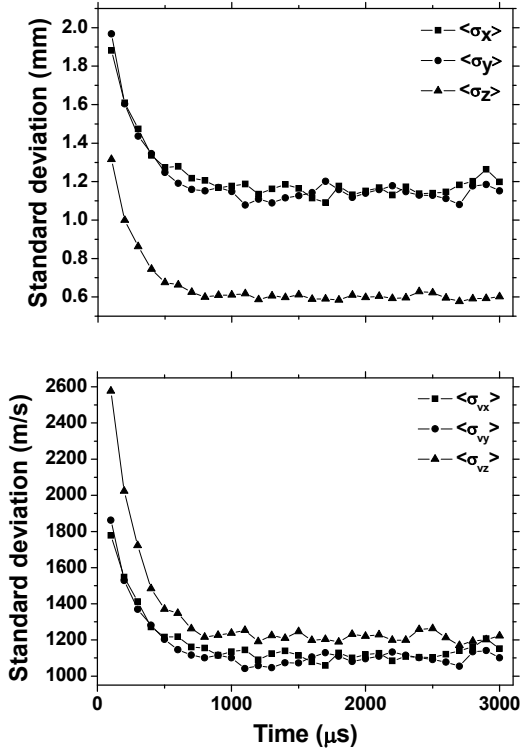}
\caption{Evolution of the mean value of the standard deviations of the positions and velocities}
\label{Fig.8}
\end{center}
\end{figure}

In reference \cite{bibi9} simulated results for the behavior of $^{6}Li^{+}$ ions in residual $H_{2}$ buffer gas are presented as well. Realistic $^{6}Li^{+}-H_{2}$ interaction potential was used. The simulations focus on the spatial and velocity distributions of the trapped $^{6}Li^{+}$. The trapping voltages were $U_{DC}=0$ and $V_{AC}=80\ V$ with a frequency of $1.3\ MHz$ and a residual  pressure of $5\times10^{-4} mbar$. In order to test our method, we used it to reproduce the results of these simulations. We have a calculation time of $100\ \mu s$ and the total time $3000\ \mu s$. In figure \ref{Fig.7} we show the evolution of $\sigma_{x}$, $\sigma_{y}$ and $\sigma_{z}$ for the five periods following a trapping time of $1000\ \mu s$. The mean values of these standard deviations as function of the trapping time are given in figure \ref{Fig.8}.

We observe a thermalization after $1000\ \mu s$. The average value of the standard deviations for the thermalized ions are $\langle\sigma_{x}\rangle=1.16\ mm$, $\langle\sigma_{y}\rangle=1.14\ mm$, $\langle\sigma_{z}\rangle=0.6\ mm$, $\langle\sigma_{v_{x}}\rangle=1121\ m/s$, $\langle\sigma_{v_{y}}\rangle=1094\ m/s$ and $\sigma_{v_{z}}=1217\ m/s$. They are close to the results of reference \cite{bibi9} which are $\langle\sigma_{x}\rangle=1.2\ mm$, $\langle\sigma_z\rangle=0.63\ mm$, $\langle\sigma_{v_{x}}\rangle=1115\ m/s$ and $\langle\sigma_{v_{z}}\rangle=1200\ m/s$.

\section{The dipolar excitation}\label{sec7}

The dipolar excitation is a common experimental method used for the determination of the ions oscillation frequencies (see for example \cite{bibi23}). It consists on the superposition of a dipolar alternating voltage of small amplitude and variable frequency (tickle) to the trapping field. When the tickle's frequency is equal to the ions oscillation frequencies, there is a resonance. Ions absorb energy and the amplitude of their motion increases. They may be lost from the trap.

The additional voltage is given by $C_{1} cos(\omega t)z$ where $C_{1}$ is a constant. It induces the additional force $-C_{1}Qcos(\omega t)$ in the $z$ direction. In the system of equations (\ref{eq13}), the equations for $x$ and $y$ remain unchanged, while in the right side of the equation for $z$ the additional term $-F_{d} cos(\omega_{rel}\xi)$ appears with  $F_{d}=\frac{C_{1}Q}{m\Omega^{2}}$ and  $\omega_{rel}=\frac{\omega}{\Omega}$. To apply the power series method, the cosine function is replaced by its Taylor expansion at each phase step
\begin{equation}\label{eq35}
cos(\omega_{rel}\xi)=\sum_{n=0}^{n_{max}}e_{n}(\xi-\xi_{0})^{n}
\end{equation}
with
\begin{eqnarray}\label{eq36}
e_{0} &=& cos(\omega_{rel}\xi_{0})\nonumber \\
e_{1} &=& -\omega_{rel}sin(\omega_{rel}\xi_{0})\nonumber \\
e_{n} &=& -\omega_{rel}^{2}\frac{e_{n-2}}{n(n-1)}\ for\ n\geq2
\end{eqnarray}
and $\xi_{0}$ being equal to $N\Delta \xi$ for the $N^{th}$ step.
This affects the recursion relation (\ref{eq23}) where the term $-\frac{F_{d}e_{n}}{(n+1)(n+2)}$ has to be added to the expression of $A^{z}_{n+2}$.

\begin{figure}[htb]
\begin{center}
\includegraphics[scale=0.65]{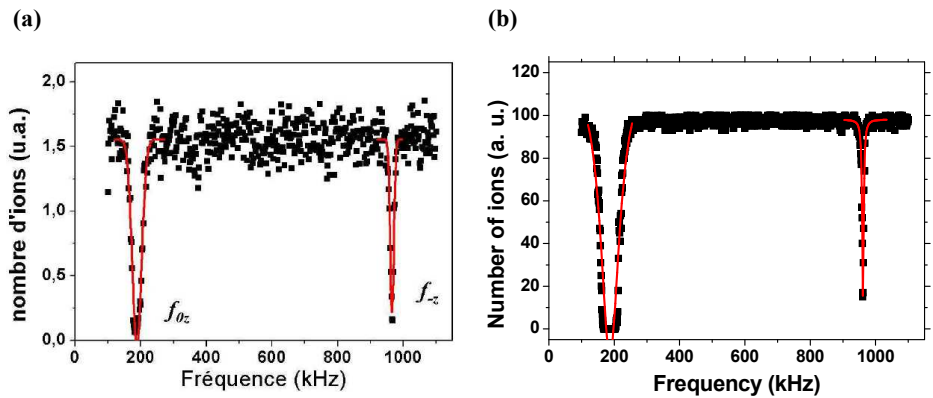}
\caption{\textbf{(a)} Experimental dipole excitation spectrum of $^{6}Li^{+}$ for $V_{AC} = 60\ V$ and $f = 1.15\ MHz$ obtained from Ref. \cite{bibi11}. \textbf{(b)} The simulated spectrum for the same conditions}
\label{Fig.9}
\end{center}
\end{figure}

In the case of the LPCTrap, the dipolar excitation has been realized by applying an additional voltage $V_{t} cos(\omega t)$ to one of the inner rings, R1, and $-V_{t} cos(\omega t)$ to the second one R2 \cite{bibi11}\cite{bibi12}. The tap is loaded with $^{6}Li^{+}$ ions, the tickle is applied for $10\ ms$ with some frequency $\omega$ and then the number of the ions remaining in the trap is counted with an MCP detector. The operation is repeated for different $\omega$. When there is no resonance, the number of the ions remaining in the trap is almost constant. At the resonance, this numbers clearly decreases as is shown in figure \ref{Fig.9}(a).

To simulate the effect of such a field, we used SIMION7 to calculate the electric potential inside the trap when $1\ V$ is applied to R1 and $-1\ V$ to R2. These additional voltages, produce in the central region of the trap, an electric potential which can be approximated by a linear function of $z$. The excitation is then mainly axial. The region where the potential has been calculated has $-5\ mm \leq z \leq 5\ mm$,$0 \leq r \leq 5\ mm$ and we used $0.01\ mm$ step for $z$ and $r$. This potential has been fitted by the function $c_{1} z$. We find
\begin{equation}\label{eq37}
c_{1}=75.39\ m^{-1}
\end{equation}
$C_{1}$ is equal to $c_{1} V_{t}$.

In order to approach the experimental conditions as well as possible we considered, besides the dipole excitation, also the cooling by buffer gas. Every time step then, the collisions of every ion with the molecules of the buffer gas were implemented by Monte Carlo simulations according to the method described in section \ref{sec6}. The buffer gas was hydrogen at $300\ K$ and $6\times 10^{-6}\ mbar$.

For the LPCTrap, the tickle was applied for $10\ ms$ after the cooling of the ions by the buffer gas. The trapped ions are $^{6}Li^{+}$. The RF has $60\ V$ amplitude and $1.15\ MHz$ frequency \cite{bibi11}. To simulate this, we start for this working point by considering the evolution of 500 ions up to $55\ ms$ under a hydrogen pressure of $6\times 10^{-6}\ mbar$. The positions and the velocities of 100 of the remaining ions were stored and used as initial conditions for the simulation of the dipole excitation with the different tickle frequencies which has a constant amplitude of $0.6\ V$. We varied the tickle's frequency with a $1\ kHz$ step and we counted the number of ions remaining in the trap after $10\ ms$ excitation. The result is shown in figure \ref{Fig.9} (b).

Fitting the resonance peaks with lorentzians, gives for the fundamental frequency $186.7\pm 0.3\ kHz$. The experiment gives $188\ kHz$. In figure \ref{Fig.9} (b), the second peak has the frequency $961.6\pm0.05\ kHz$. The sum of the frequencies of the two peaks, i.e. $1148.3\pm 0.3\ kHz$, is almost equal to the frequency of the trapping field.

\section{Conclusion}

We developed a method for the simulation of the dynamics of ions trapped in a Paul trap with the presence of terms in the potential up to the order 10. The equations of motions of the ions are solved numerically. Every time step, the coordinates of the ion are represented by a 15 degree polynomials of time. The coefficients of the polynomials are determined by imposing the continuity of the position and the velocity and by using the power series method. When applied to particular cases, the method gives oscillation amplitudes and frequencies close to those obtained by SIMION7, however our calculations are much faster.

The method was then applied to an existing trap. When knowing the coefficients of the trapping potential, we could consider the evolution of the position of hundreds of ions up to several tens of milliseconds. This is the time scale for the trap. By scanning the Mathieu parameters and counting the number of ions remaining in the trap after 2 ms for every working point, we get the first stability domain which shows the nonlinear resonances.

The buffer gas cooling is introduced by a Monte Carlo method where The hard sphere model is used, here again the results are close to the experimental ones and those obtained by simulations.

At the end we introduced a dipole excitation to ions cooled by the buffer gas. The results show good agreement with the experiment.

\section*{Aknowledgements} The present work has been accomplished under the project No 09-ADV826-07 funded by KACST (King Abdul Aziz City for Science and Technology) through the Long Term Comprehensive National Plan for Science, Technology and Innovation program in Saudi Arabia. The authors thank Yvan Merrer from LPC Caen for the technical informations about their trap.

\end{document}